\documentclass[twocolumn,showpacs,preprintnumbers,amsmath,amssymb]{revtex4}

\usepackage{epsfig}

\begin{document}


\title{Stability and structure of two coupled boson systems in an 
external field}

\author{T. Sogo}
\author{O. S{\o}rensen}
\author{A.S. Jensen}
\author{D.V. Fedorov}
\affiliation{ Department of Physics and Astronomy,
        University of Aarhus, DK-8000 Aarhus C, Denmark }

\date{\today}

\begin{abstract}
The lowest adiabatic potential expressed in hyperspherical coordinates
is estimated for two boson systems in an external harmonic
trap. Corresponding conditions for stability are investigated and the
related structures are extracted for zero-range interactions.  Strong
repulsion between non-identical particles leads to two new features,
respectively when identical particles attract or repel each other.
For repulsion new stable structures arise with displaced center of
masses.  For attraction the mean-field stability region is restricted
due to motion of the center of masses.
\end{abstract}


\maketitle

\paragraph*{Introduction.}

Bose-Einstein condensation is by now routinely realized in many
laboratories \cite{pet01,pit03}. Static and dynamic properties are
currently investigated by use of Feshbach resonances to manipulate the
effective two-body interactions. Two-component and mixed systems open
for a variety of combinations.  Two-component condensed bosons systems
consisting of the same atoms in different spin states are formed
\cite{MBG97,HME98} and collective oscillations studied for $^{87}$Rb
\cite{MMF00}.  Condensates of two different species were recently
created combining $^{87}$Rb and $^{41}$K \cite{MMR02}.  Also mixed
systems of condensed boson and degenerate fermions was recently
experimentally obtained \cite{tru01}.

Theoretical descriptions often use the mean-field approximation with
zero range interactions \cite{pet01,pit03}. Both stability and average
properties are successfully described for one-component boson systems
which so far are the most intensively studied.  The simplest model for
two-component and mixed systems is a coupled mean-field approximation
\cite{BCP97,Oh98,EG99}.  Two-cluster features may then be accounted for by
using mean-field intrinsic and relative coordinates.  However, for
such systems a more complicated cluster structure may be favorable
corresponding to lower energy.  Competing or maybe co-existing new
types of structures may appear. The stability conditions may be
altered when degrees of freedom beyond the mean field are allowed.

The mean-field models are unable to investigate effects of correlations
\cite{sor03}. Correlations can be studied 
in hyperspherical models, 
which have been applied to one-component symmetric boson systems in
external fields \cite{sor02}.  An extension to two-component systems
requires at least introduction of a relative and a size coordinate for
each subsystem. Compared to a one-component system the difficulties
increase substantially. Therefore, before embarking on such a large
project, it is advisable and illuminating to study the decisive
large-distance limit with zero-range interactions and simple
approximate wave functions.  The purpose of the present letter is to
provide estimates for conditions of stability and occurrence of new
structures for two coupled boson systems.

\paragraph*{Theoretical method.}

The $N$ bosons have masses $m_i$ and coordinates $\vec r_i$. We assume
a division into two systems, $A$ and $B$, of identical bosons with
particle numbers $N_A$ and $N_B$ ($N = N_A + N_B$) and total masses
$M_A \equiv N_A m_A $ and $M_B \equiv N_B m_B $ ($M \equiv M_A +
M_B$).  We shall use hyperspherical coordinates with the hyperradius
defined by \cite{sor02,boh98}
\begin{equation} \label{e10}
 m \rho^2 \equiv \frac{1}{M} \sum_{i<j}^N m_i m_j r_{ij}^2 =
 \sum_{i=1}^N m_i r_i^2 -  M R^2 \;,
\end{equation}
where $m$ is an arbitrary normalization mass, $\vec r_{ij} = \vec r_i
- \vec r_j$ and $\vec R = \sum_{i} m_i \vec r_i /M$ is the
center-of-mass coordinate. The remaining degrees of freedom are
described by the hyperangles collectively denoted by $\Omega$.
A system divided into two different subsystems of non-identical
particles requires independent parameters for each component.  The
size coordinates defined as the corresponding two hyperradii are then
the natural choice analogous to mean-field coordinates for each
component. In addition we also allow the correlations corresponding to
relative motion of the two center of masses. This is the minimum
number of parameters for a two-component system where the sizes and
center of mass positions can be independently varied.  
Thus we define the hyperradii and center-of-mass
coordinates for the two systems $A$ and $B$
\begin{eqnarray} \label{e15}
  \rho_A^2 \equiv \frac{1}{N_A} \sum_{i<j}^{N_A} r_{ij}^2 \; , \;\;
  \rho_B^2 \equiv \frac{1}{N_B} \sum_{N_A<i<j}^{N} r_{ij}^2 \; ,& \\ 
\label{e20}
 \vec R_A = \frac{1}{N_A}\sum_{i=1}^{N_A}  \vec r_i   \; , \;\;
 \vec R_B = \frac{1}{N_B} \sum_{i=N_A+1}^{N} \vec r_i  \; .& 
\end{eqnarray}
The average separation of the two components is then given by 
the coordinate $\vec r \equiv  \vec R_A  - \vec R_B$.
The three length coordinates, $\rho_A$, $\rho_B$, and $r$, are related
to $\rho$ by
\begin{equation} \label{e25}
 m \rho^2 =  m_A \rho_A^2 +  m_B \rho_B^2 +  m_r r^2 \;,\;\; 
 m_r \equiv \frac{M_A M_B}{M_A + M_B} \;  .
\end{equation}

We aim at using the hyperspherical adiabatic expansion method.  We
only include the first adiabatic angular wave function, i.e.
\begin{equation} \label{e30}
 \Psi = \rho_A^{2-3N_A/2} \rho_B^{2-3N_B/2} r^{-1} f(\rho_A,\rho_B,r) 
 \Phi(\Omega) \; ,
\end{equation}
where the separable center-of-mass motion is removed. We
first assume that $s$-waves dominate in the angular expansion and
second that $\Phi$ is independent of the angles. Then the wave
function is automatically symmetric under permutations of particles
within each subsystem.  These assumptions are severe, but not
unreasonable for spatially extended dilute systems.

The three short-range interactions are all chosen as
$\delta$-functions with strengths expressed in terms of the
corresponding two-body scattering lengths $a_A$, $a_B$, and $a_{AB}$,
i.e.
\begin{eqnarray} \label{e35}
&&V_{A}=\sum_{i<j=1}^{N_A}g_{A}\delta^{(3)}(\vec r_{ij}) \; ,\;\;
g_{A}=\frac{4\pi\hbar^2a_{A}}{m_A} \;, \\
&&V_{B}=\sum_{N_A<i<j=1}^{N}g_{B}\delta^{(3)}(\vec r_{ij}) \;, \;\;
g_{B}=\frac{4\pi\hbar^2a_{B}}{m_B} \;, \\
&&V_{AB}=\sum_{i=1}^{N_A}\sum_{j=N_A+1}^{N}g_{AB}\delta^{(3)} (\vec r_{ij}) \;,\;\;
g_{AB}=\frac{2\pi\hbar^2a_{AB}}{\mu_{AB}} \; ,
\end{eqnarray}
where $\mu_{AB}\equiv m_A m_B/(m_A+m_B)$.  Integrating these
interactions over all hyperangles we arrive at the Schr\"odinger
equation which determines $f$ in Eq.~(\ref{e30}), i.e.
\begin{eqnarray} \label{e55}
 \Big(-\frac{\hbar^2}{2 m_A}  \frac{\partial^2}{\partial \rho_A^2} -
  \frac{\hbar^2}{2 m_B} \frac{\partial^2}{\partial \rho_B^2} &-&
 \frac{\hbar^2}{2 m_r} \frac{\partial^2}{\partial r^2}  \;, \;\; \\ \nonumber
 +  U_{eff} &-& E\Big) f(\rho_A,\rho_B,r)= 0  \;, 
\end{eqnarray}
where the adiabatic effective potential for $N_A \gg 1$ and $N_B \gg
1$ is
\begin{eqnarray}
\frac{2 m_A}{\hbar^2} U_{eff}(\rho_A,\rho_B,r) =  \frac{\rho_A^2}{ b_t^4} +
  \frac{m_B\rho_B^2}{m_Ab_t^4}  +  \frac{m_r r^2}{m_A b_t^4}   
   \;\; \nonumber \\  \label{e60} 
 + \frac{9 N_A^2}{4 \rho_A^2}   +  \frac{9 m_A N_B^2}{4m_B \rho_B^2}  
 + \frac{3^{3/2}}{2\pi^{1/2}} \Big(\frac{N_A^{7/2}a_{A}}{\rho_A^3} +
 \frac{m_Aa_{B}N_B^{7/2}}{m_B\rho_B^3} \;\; \\ \nonumber 
 + \frac{m_A(N_A N_B)^{7/4}a_{AB}}{\mu_{AB}(\rho_A \rho_B)^{3/2}}  
 I(\rho_A,\rho_B,r)\Big) \;, \;\;
\end{eqnarray}
where we have defined the trap length for the external-field frequency
$\omega$ as $b_t^2 \equiv \hbar/(m_A \omega)$.  The dimensionless
function $I$ depends on $N_A$, $N_B$, $\rho_A /r$, and $\rho_B /r$. It
arises from the angular average of the interaction $V_{AB}$ between
the particles in systems $A$ and $B$.

\paragraph*{Effective potential.}

The effective potential $U_{eff}$ in Eq.~(\ref{e60}) consists of three
types of terms, i.e. from the external field (first three), the
generalized centrifugal barrier (next two), and the angular average of
the interactions and kinetic energies (last three). In total,
$U_{eff}/(\hbar\omega)$ depends on the particle numbers $N_A$ and
$N_B$, the ratio $m_A/m_B$, three coordinates and three scattering
lengths measured in units of $b_t$.

The function $I$ in Eq.~(\ref{e60}) is a double integral with rather
simple properties, i.e.it is monotonous as function of $r$, it
vanishes when $r > \rho_A + \rho_B$, has a positive maximum and
vanishing first derivative at $r=0$, and it is unity when $N_A = N_B$ and
$\rho_A = \rho_B$.

The behavior of $U_{eff}$ is qualitatively different for attractive
and repulsive two-body interactions. The interaction terms
($\rho^{-3}$) dominate for small distances, and for strong attraction
$U_{eff}$ has no minimum.  The system is always spatially confined by the
external harmonic field. When subsystems $A$ and $B$ are identical, the
effective potential has a local minimum when $N|a|/b_t < 0.67$
\cite{pet01,pit03}. For asymmetric systems the corresponding stability
conditions are more complicated, as we shall see.

\begin{figure}[hbt]
\vspace*{-0mm}
\begin{center}
\epsfig{file=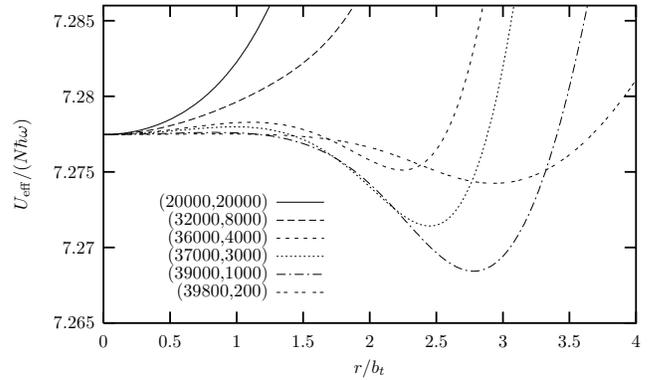,width=8.4cm,angle=0}
\end{center}
\vspace*{-0.5cm}
\caption[]{The effective potential in Eq.~(\ref{e60}) as a function of 
$r/b_t$ for $^{87}$Rb masses $m_A = m_B = 1.44 \times 10^{-25}$kg, trap
length $b_t = 2.22 \mu$m ($\omega = 2\pi \times 23.5$ Hz), particle numbers
$N=N_A+N_B=40000$, and $a_A = a_B = a_{AB} = 103 a_0$, where $a_0$ is
the Bohr radius. The curves correspond to the different values of
$(N_A,N_B)$ given on the figure.}
\label{fig1}
\end{figure}

To illustrate, we show in Fig.~\ref{fig1} for three repulsive
scattering lengths the minimum of $U_{eff}$ with respect to $\rho_A$
and $\rho_B$ as a function of $r$ and various divisions of the total
particle number 40000.  For symmetric divisions coinciding centers of
mass ($r=0$) are preferred. As the asymmetry increases ($N_B < 0.134 N_A$), a
deeper minimum develops at finite $r$-values comparable to the trap
length.  The minimum at $r=0$ remains stable because the barrier
separating the minima is much higher than the zero-point motion in
both wells. The deepest minimum occurs for the asymmetric system
with $N_B/N_A \approx 1/40$.

\paragraph*{Structure.}

The basic characteristics of a system are size and energy. The
root-mean-square distance $r_A$ is the measure of the size of system
$A$ defined by
\begin{eqnarray} \label{e68}
  N_A \bar r_A^2 \equiv
 \langle \sum_{i}^{N_A} (\vec r_{i} -\vec R_{A})^2 \rangle
 = \langle \rho_A^2 \rangle\; ,
\end{eqnarray}
where $\langle \rangle$ denotes an expectation value.  Correspondingly we
define mean distances $\bar r$ and  $\bar r_B$ for the total and the
$B$-system, i.e.  $N \bar r^2 = \langle \rho^2 \rangle$ and $N_B \bar
r_B^2 = \langle \rho_B^2 \rangle$.

When the different particles attract each other, i.e.  $a_{AB} < 0$,
the effective potential only has the minimum for coinciding centers
where $r=0$. The structures are then similar to those of one-component
systems. Therefore, we shall here concentrate on the different
structures appearing for repulsive cases where $a_{AB} > 0$.

\begin{figure}[hbt]
\vspace*{-0mm}
\begin{center}
\epsfig{file=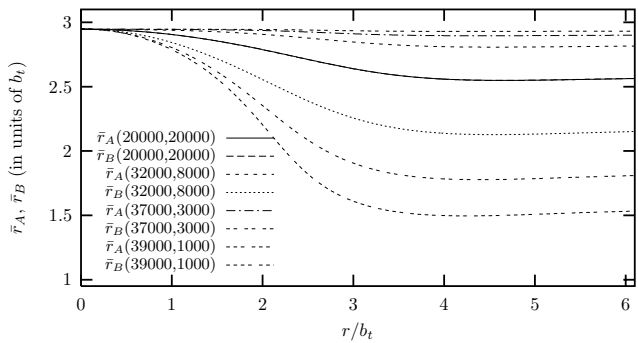,width=8.4cm,angle=0}
\end{center}
\vspace*{-0.5cm}
\caption[]{The root-mean-square radii $\bar r_A$ and $\bar r_B$ as 
a function of $r$ for the systems specified in Fig.~\ref{fig1}.}
\label{fig2}
\end{figure}

The average sizes corresponding to Fig.~\ref{fig1} are shown in
Fig.~\ref{fig2}.  For the minimum at $r=0$ we obtain large sizes,
i.e. $\bar r_A \approx \bar r_B \approx \bar r \approx 3 b_t$. This
structure again closely resembles the one-component system.  For a
given division $N_B/N_A$ both subsystems contract and stabilize on a
lower level as $r$ increases to a few time times the trap length.  The
smaller the system, the larger the contraction.

The effect of the repulsion is that the two systems try to avoid each
other while staying at relatively small distances as dictated by the
external field. The large subsystem ($A$) is exploiting the space
almost like it was alone in the trap. When one subsystem ($B$) is small,
it becomes advantageous to place about half of its particles outside
the other subsystem. Then the repulsion on $B$ from the trap and the
subsystem $A$ is minimized.

The minima of the effective potentials define the preferred structure.
As $N_B/N_A$ changes, the system with the largest particle number
($N_A$) always has the same size of about $3b_t$ in these minima. In
contrast, the size of the other subsystem decreases after $N_B < 0.134
N_A$ where the lower minimum at finite $r$ appears. The values of
$\bar r_B$ converge towards $b_t$ when $N_B/N_A \rightarrow 0$.

The center of mass of the small system (B) remains within the radius
$\bar r_A$ until $N_B \approx 0.004 N_A$. For even smaller subsystems
the center of $B$  moves a little outside $\bar r_A$, but still the
distance between the centers at the minimum remains smaller than $\bar
r_A + \bar r_B$. Therefore, complete separation between the systems
does not occur.

\paragraph*{Stability.}

\begin{figure}[hbt]
\vspace*{-0mm}
\begin{center}
\epsfig{file=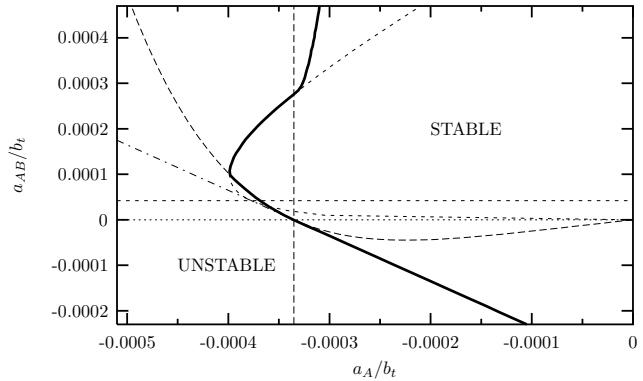,width=8.4cm,angle=0}
\end{center}
\vspace*{-0.5cm}
\caption[]{Stability regions for a two-component system with 
$N_B = N_A = 2000$ as a function of $a_A = a_B < 0$ and $a_{AB}$. The
thick solid line separates stable (right side) from unstable (left
side) regions when we allow three degrees of freedom, i.e. $\rho_A$,
$\rho_B$ and $r$. The different curves are related to Eqs.~(\ref{e70})
(dot-dashed), (\ref{e80}) and (\ref{e85}) (long-dashed), (\ref{e90})
and (\ref{e95}) (short-dashed), respectively.}
\label{fig3}
\end{figure}

The stability of the system is guaranteed for repulsion within both
subsystems, i.e. when $a_A>0$ and $a_B>0$. Instability arises for
attraction where one or both subsystems can collapse into high-density
states of lower energy. In Fig.~\ref{fig3} we illustrate these
stability regions for a specific system as a function of the
scattering lengths $a_A = a_B < 0$ and $a_{AB}$. The results are
extracted from the effective potential in Eq.~(\ref{e60}). Stability
is defined as the existence of a local minimum.  We first consider
$r=0$ and search for minima by variation of $\rho_{A}$ and
$\rho_{B}$. No stationary point exists for $N_A = N_B$ when
\begin{equation} \label{e70}
\frac{N_A(a_{A}+a_{AB})}{b_t}
\leq - \frac{2^{3/2}\pi^{1/2}}{5^{5/4}} \approx -0.67 \; ,
\end{equation}
which follows the critical stability line in Fig.~\ref{fig3} for
negative $a_{AB}$-values.

Maintaining $r=0$ we find that any stationary point of $U_{eff}$ has
negative curvature (second derivative) in one of the principal
directions when
\begin{eqnarray} \label{e80}
\frac{a_{AB}}{b_t} &\leq& 
-\frac{5^5N_A^4}{2^8\pi^2}
\left(\frac{a_{A}}{b_t}\right)^5
+\frac{1}{4}\frac{a_{A}}{b_t}  \; , \\ \label{e85}
\frac{N_Aa_{A}}{b_t} &\leq& - \frac{2^{3/2}\pi^{1/2}}{5^{5/4}} 
\approx - 0.67 \; ,
\end{eqnarray}
where Eq.~(\ref{e80}) for $0 < a_{AB}/b_t \lesssim 0.0001$ follows the
critical stability curve in Fig.~\ref{fig3}.  Eqs.~(\ref{e70}),
(\ref{e80}), and (\ref{e85}) are comparable with the mean field
approximation \cite{BCP97}.

When $0.0001 \lesssim a_{AB}/b_t$, the minimum of $U_{eff}$ occurs for
finite $r$, i.e. separation of the center of masses is favorable.
Furthermore, the curvature at $r=0$ in the $r$-direction is negative
when
\begin{eqnarray} \label{e90}
\frac{a_{A}}{b_t} &\leq& 
 \frac{a_{AB}}{b_t}
-\frac{\pi^{2/5}}{2^{1/5}N_A^{4/5}}
\left(\frac{a_{AB}}{b_t}\right)^{1/5} \; , \\  \label{e95}
\frac{N_Aa_{AB}}{b_t} &\geq& \frac{\pi^{1/2}}{2^{3/2}5^{5/4}} \approx 0.084\; ,
\end{eqnarray}
where Eq.~(\ref{e90}) is decisive by defining the critical stability
curve in Fig.~\ref{fig3} for $0.0001 \lesssim a_{AB}/b_t \lesssim
0.00028$.  Eq.~(\ref{e95}) resembles the mean-field expression derived
in \cite{Oh98}.

\begin{figure}[hbt]
\vspace*{-0mm}
\begin{center}
\epsfig{file=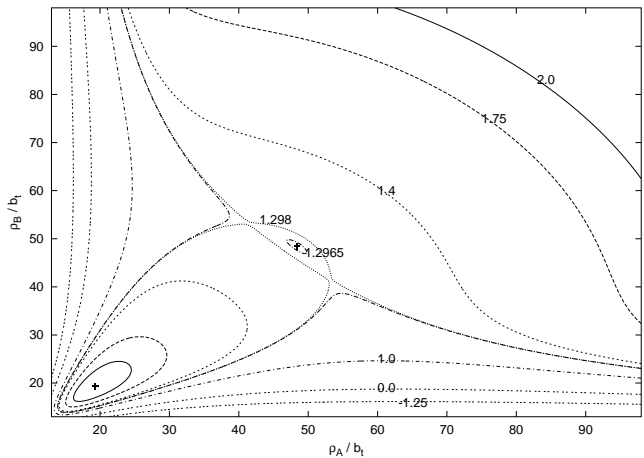,width=8.8cm,angle=0}
\end{center}
\vspace*{-0.5cm}
\caption[]{Contour plot of the potential $U_{eff}/(N\hbar\omega)$
in Eq.~(\ref{e60}) as a function of $\rho_A$ and $\rho_B$ for $r=0$,
$N_A = N_B = N/2 =2000$, $a_{AB}=0.0002 b_t$, and $a_{A}=a_{B}=-0.00042 b_t$. }
\label{fig4}
\end{figure}

The critical stability curve is in agreement with mean-field
computations for $a_{AB}/b_t \lesssim 0.0001$, but strongly deviating
for $0.0001 \lesssim a_{AB}/b_t$ where the system exploits the
$r$-degree of freedom and becomes unstable for a range of attractive
$a_{A}/b_t$-values. This is illustrated in Fig.~\ref{fig4} for a set
of scattering lengths where a very shallow local minimum is present
for coinciding centers, i.e. $r=0$.  Along the line defined by $\rho_A
= \rho_B$ a maximum is found around $20 b_t$ and a minimum at $48.4
b_t$. This degree of freedom is analogous to the one-component
mean-field coordinate.
The two-component mean-field coordinates are presumably similar to the
use of $\rho_A$ and $\rho_B$ for $r=0$. A more detailed comparison is
required to distinguish.

Moving now in the direction where the sum of $\rho_A$ and $\rho_B$
is constant and the difference varies, we find rather small
barriers for $(\rho_A, \rho_B) \approx (40 b_t,50b_t), (50
b_t,40b_t)$.  This degree of freedom is analogous to the two-component
mean-field coordinates. We now decrease the repulsive scattering
length $a_{AB}$ by a factor of two from the value in Fig.~\ref{fig4}
while maintaining $a_{A}$.  Then the shallow minimum disappears in
agreement with the Fig.~\ref{fig3} where the decrease of $a_{AB}$
corresponds to crossing the curve described in Eq.~(\ref{e80}) where
one curvature changes sign.  The larger repulsion in Fig.~\ref{fig4}
stabilizes the system as predicted in the mean-field approximation.
However, if the $r$-degree of freedom also is allowed, the system
collapses by moving the centers apart from each other as described by
Eq.~(\ref{e90}).  The local minimum in Fig.~\ref{fig4} is in fact
unstable.  Stability is regained for a finite $r$-value when $a_A$ is
less attractive and $a_{AB}> 0.00028 b_t$, see the upper middle part
of Fig.~\ref{fig3}.

The critical line for stability of each of the independent systems $A$
and $B$ is also shown in Fig.~\ref{fig3} as $a_A/b_t = -0.67/N_A =
-0.00034$. The repulsion between $A$ and $B$ stabilizes the total
system even when each component would collapse when left alone. The
reason is that the contraction of each subsystem increases the mutual
repulsion. The balance of these forces results in increased stability
of the total system for moderate repulsion.  In contrast, less
stability results for strong repulsion where the collapse proceeds by
circumventing the individual barriers via relative center-of-mass
motion.

The two minima in Fig.~\ref{fig1} are not found when 
$a_{A}$ = $a_{B}$ and $N_A=N_B$.  This feature only seems to
occur for asymmetric systems.
When both minima are present, each corresponds to a stable
configuration.  The vibrational zero-point energy is much smaller than
the hardly visible, but stabilizing barrier in Fig.~\ref{fig1}.
Furthermore, estimates of the WKB tunneling probability show decay
rates much smaller than the frequency of the external harmonic field.
Co-existing structures are then possible.

\paragraph*{Conclusion.}

We have investigated coupled two-component boson systems in a
confining trap. The degrees of freedom are the sizes of the two
components and the center-of-mass distance. Two new features
emerge. First, new stable structures appear in the trap when identical
and different pairs of bosons all repel each other and one of the
components  at the same time has many more particles than the other.
Then a finite distance between the centers of the two components is
energetically favored and two stable co-existing structures are
possible. 

The second new feature appears when identical bosons attract
each other and different bosons repel each other. Some structures
previously computed to be stable in the mean-field approximation are
unstable. In the present model with more degrees of freedom, when the
repulsion is sufficiently strong, the mean-field minimum becomes
unstable against relative motion of the distance between the two
centers of masses. The mean-field regions of stability are then
further restricted.

The inaccuracies in the computations are all related to the choice of
parameter space. Improving by including two-body correlations is
possible for example by increasing the variational space of the wave
function as done in \cite{sor03,sor02} for a one-component system.
These improvements are then correspondingly expected to be substantial
for hyperradii down to the range of the interaction, relatively small
for hyperradii corresponding to condensate distances, and
insignificant for larger distances. Thus the qualitative structure
would remain unchanged and at the crucial distances for condensate
formation the results are rather close to being quantitatively
correct.

\end{document}